\documentclass[aps,pre,twocolumn,superscriptaddress,floatfix,longbibliography]{revtex4-2}

\usepackage{amsmath,amssymb}
\usepackage{graphicx}
\usepackage{booktabs}
\usepackage[colorlinks=true,linkcolor=blue,citecolor=blue,urlcolor=blue]{hyperref}

\newcommand{\ave}[1]{\langle #1 \rangle}
\newcommand{\kb}{\bar{k}}
\newcommand{\cb}{\bar{c}}
\newcommand{\vecA}{\mathrm{vec}^2 A}
\newcommand{\code}[1]{\texttt{#1}}

\begin{document}

\title{The giant component of complex hypergraphs:\\
automated generating function calculations}

\author{Alexei Vazquez}
\email{alexei@nodeslinks.com}
\affiliation{Nodes \& Links Ltd, Salisbury House, Station Road, Cambridge, CB1 2LA, UK}

\begin{abstract}
Complex hypergraphs (chygraphs) contain graphs, hypergraphs, multiplex and
interacting networks as special cases, and the percolation threshold of all of
them follows from one symbolic calculation: the spectrum of a tensor $A$ built
from four matrices of first moments. That calculation stops at the threshold.
Here I show that $A$ is the Jacobian, at its trivial fixed point, of a
non-linear self-consistency map whose non-trivial fixed point is the giant
component fraction, so threshold and order parameter are two orders of one
object on one index set. Carrying the expansion one order further gives the
critical amplitude $B$ in $S=B\Lambda+O(\Lambda^2)$ in closed form, and exposes
a hierarchy: first moments fix the threshold, second moments the amplitude, and
only the generating functions themselves the order parameter away from it.
Dropping the assumption that a complex's participation in different layers is
independent generalises the tensor again, replacing unconditional first moments
by inclusion-biased second moments: distributions with identical marginals have
different thresholds and different order parameters. Six constructions from the
literature are then solved by substitution, among them AND- against OR-logic
hypergraph percolation, which share a chygraph and differ in one generating
function, and SIR epidemics with two levels of mixing, whose household
reproduction number falls out of the tensor. A reported failure of mean-field
theory on strongly clustered graphs is shown to be a failure of applying it to
the wrong object: mapped onto the treelike backbone rather than the clustered
graph, the same calculation reproduces the exact threshold, order parameter and
critical exponents. Everything is implemented in one class, in which a chygraph
is specified once by its generating functions and then returns all three, and
validated against Monte Carlo simulation.
\end{abstract}

\maketitle

\section{Introduction}

Percolation is the reference problem of network science, and the generating
function formalism is its reference tool \cite{molloy1998,callaway2000}. The
proliferation of higher order networks --- hypergraphs, simplicial complexes,
multiplex, interacting and interdependent networks
\cite{kivela2014,battiston2020,battiston2021,bianconi2021} --- has multiplied
the number of independent generating function calculations, each cumbersome and
each hard to check.

To consolidate that work I introduced complex hypergraphs, or chygraphs, a set
of complexes where complexes are hypergraphs with vertex sets in the set of
complexes \cite{vazquez2023pre}, and obtained the emergence of the giant
component from a generating function calculation on that construction. In a
follow-up \cite{vazquez2024comnet} I showed that once a higher order network is
mapped to a chygraph, its percolation threshold is obtained by substituting four
matrices into a single symbolic calculation, reproducing results for graphs,
hypergraphs, multiplex hypergraphs, network motifs and interacting networks that
had been derived separately over two decades.

That programme delivers the threshold and stops there. The quantity an
application usually wants is the order parameter: the fraction $S$ of nodes in
the giant component, not just whether it is zero. This work closes that gap.

The starting observation is structural. The tensor $A$ of
Refs.~\cite{vazquez2023pre,vazquez2024comnet}, whose entries are the first
moments $\ave{\kappa}$, $\ave{\bar\kappa}$, $\ave{s}$, $\ave{\bar s}$, is not an
object in its own right: it is the Jacobian of a non-linear map at its trivial
fixed point. The threshold diagnostics $\theta=-\det(\vecA)$ and
$\Lambda=\max\mathrm{eig}(-\vecA)$ are properties of the linearisation; the
order parameter is the non-trivial fixed point of the map itself. Recovering the
map is therefore not a new formalism but a step backwards from a derivative to
the function it came from.

Two things follow. First, $S$ requires strictly more input than $\theta$:
generating functions rather than their first derivatives. I make the obstruction
precise --- two chygraphs with identical $\ave{\kappa},\ave{\bar\kappa},
\ave{s},\ave{\bar s}$, hence identical $\theta$ and $\Lambda$, can have
different $S$ --- and show that the extra input is already implicit in every
example treated in Ref.~\cite{vazquez2024comnet}. Second, and more useful, the
leading behaviour of $S$ near the threshold does \emph{not} need the full
generating functions. Expanding the map one order beyond the Jacobian gives
\begin{equation}
  S = B\,\Lambda + O(\Lambda^2)
\end{equation}
with $B$ in closed form in terms of second derivatives of the same generating
functions whose first derivatives build $A$. The result is a three-tier
hierarchy of what each level of moment data buys, summarised in
Table~\ref{tab:hierarchy}.

A third result comes from asking what the recovered map assumes. The published
tensor treats a complex's participation in different layers as independent.
Once the map is written with joint generating functions that assumption becomes
visible and removable, and the tensor generalises to carry inclusion-biased
second moments in the slots where it carried unconditional first moments. The
correction is not small: three joint degree distributions with identical
marginals give three different thresholds.

Section~\ref{sec:constructions} puts all of this to work on constructions that
have been solved separately in the literature. Each is handled by writing down
its chygraph mapping and substituting, which is the point of the programme: the
calculation is done once, in Secs.~\ref{sec:map}--\ref{sec:joint}, and reused.

The deliverable of that programme is a piece of software, so the exposition
follows the order in which the software works rather than the order in which the
theory was originally built. Section~\ref{sec:chygraphs} fixes the chygraph and
the four moment matrices; Sec.~\ref{sec:map} writes the non-linear map, solves
it, and obtains the giant component; Sec.~\ref{sec:threshold} linearises the map
and recovers the published threshold tensor as its Jacobian; and
Sec.~\ref{sec:amplitude} carries the expansion one order further. Every quantity
reported below is named with the method that computes it, collected in
Table~\ref{tab:code}, and every construction is shown as it is actually supplied
to the code in Fig.~\ref{fig:code}.

\begin{table}[b]
\caption{\label{tab:hierarchy}What each level of input determines.}
\begin{ruledtabular}
\begin{tabular}{ll}
quantity & input required \\
\colrule
threshold $\theta$, $\Lambda$ & first moments \\
critical amplitude $B$        & $+$ second moments \\
$S$ away from the threshold   & the generating functions \\
\end{tabular}
\end{ruledtabular}
\end{table}

\section{Chygraphs}\label{sec:chygraphs}

A chygraph is a set of complexes where complexes are hypergraphs with a vertex
set in the set of complexes \cite{vazquez2023pre}. Complex $A$ \emph{includes}
complex $B$ when $B$ is in the vertex set of $A$; equivalently $B$ is
\emph{included in} $A$. An \emph{atom} is a complex that includes no complexes,
and layers $l=0,\dots,L-1$ set apart complexes with different statistical
properties. Graphs, hypergraphs, multiplex hypergraphs, network motifs and
interacting hypergraphs all map onto this construction; the mappings are given
in Ref.~\cite{vazquez2024comnet} and are used unchanged here.

Four $L\times L$ matrices summarise a chygraph at the level of first moments.
$\ave{\kappa}_{lk}$ is the expected number of layer-$k$ complexes that include a
layer-$l$ complex, $\ave{s}_{lk}$ the expected number of layer-$k$ complexes in
the intra-complex component of a layer-$l$ complex, and
$\ave{\bar\kappa}_{lk}$, $\ave{\bar s}_{lk}$ are the corresponding excess
averages. These four are the whole input of the published threshold calculation.
The order of what follows is deliberate: Sec.~\ref{sec:map} builds the
non-linear map that gives the giant component and solves it, and only then does
Sec.~\ref{sec:threshold} linearise that map and recover the published tensor
from it, which is the relation the rest of the paper rests on.

\section{The giant component}\label{sec:map}

\subsection{Generating functions}

Let $\Phi^l_k(x)$ be the probability generating function of the number of
layer-$k$ complexes that include a layer-$l$ complex, and $G^l_k(y)$ that of the
number of layer-$k$ complexes in the intra-complex component of a layer-$l$
complex, reached at a given vertex. Let $\bar\Phi^l_k$ and $\bar G^l_k$ be the
corresponding excess generating functions. By construction
\begin{align}
  \Phi^{l\prime}_k(1)&=\ave{\kappa}_{lk}, &
  \bar\Phi^{l\prime}_k(1)&=\ave{\bar\kappa}_{lk},\nonumber\\
  G^{l\prime}_k(1)&=\ave{s}_{lk}, &
  \bar G^{l\prime}_k(1)&=\ave{\bar s}_{lk},
  \label{eq:firstderiv}
\end{align}
so the four moment matrices of Sec.~\ref{sec:chygraphs} are exactly the first
derivatives of these functions at $1$; that is the whole of the relation worked
out in Sec.~\ref{sec:threshold}. In this section I take the generating functions to factorise over target
layers, which covers every mapping in Ref.~\cite{vazquez2024comnet}.

\subsection{The map}

Let $Q^{ml}_i$ be the probability that a layer-$l$ complex, reached from a
layer-$m$ complex, is not connected to the giant component through any route
other than the one used to reach it. The index $i=-$ means the complex was
reached from below, from a complex it includes, and $i=+$ from above, from a
complex that includes it. Following the two ways of leaving a complex, upward
through its chy-degree and downward through its intra-complex hypergraph, and
noting that the step used to arrive is excluded only when the return layer
equals the layer of origin, the self-consistency conditions of
Ref.~\cite{vazquez2023pre} read
\begin{align}
  Q^{ml}_- &= \prod_k \Phi^l_k\!\left(Q^{lk}_-\right)
              \prod_k \Big[\bar G^l_k\Big]_{k=m}\!\!\left(Q^{lk}_+\right),
  \label{eq:Qm}\\
  Q^{ml}_+ &= \prod_k \Big[\bar\Phi^l_k\Big]_{k=m}\!\!\left(Q^{lk}_-\right)
              \prod_k G^l_k\!\left(Q^{lk}_+\right),
  \label{eq:Qp}
\end{align}
where $[\bar F]_{k=m}$ denotes the excess generating function when $k=m$ and the
unbiased one otherwise. The index $i$ of the state reached is fixed by the
direction of the step: an upward step through $\ave{\kappa}$ lands in a
$-$ state, a downward step through $\ave{s}$ in a $+$ state. Writing $F$ for
the map defined by Eqs.~\eqref{eq:Qm}--\eqref{eq:Qp} on the $2L^2$ unknowns
$Q^{ml}_i$, indexed as $a=iL^2+mL+l$ to match the vectorisation of $A$, the
probability that a randomly chosen layer-$l$ complex is not in the giant
component is
\begin{equation}
  P^l = \prod_k \Phi^l_k\!\left(Q^{lk}_-\right)\prod_k G^l_k\!\left(Q^{lk}_+\right),
  \label{eq:P}
\end{equation}
and the giant component fraction of layer-$l$ complexes is $S_l=1-P^l$. The
fraction of nodes is $S\equiv S_0$.

\subsection{Occupation probabilities}

Occupation probabilities enter by thinning the generating functions: a node
present with probability $p$ gives $\Phi\to1-p+p\Phi$. This reproduces the
convention $\ave{\kappa}_{0l}=p\ave{k}$ of Ref.~\cite{vazquez2024comnet} at the
level of first derivatives, and makes $1-P^0$ the fraction of \emph{all} nodes
rather than of surviving nodes, so that $S\le p$ automatically. At the level of
$\theta$ only the product of occupation probabilities is visible, and their
placement among the matrices is immaterial; for $S$ it is not.

Thinning is not always available. When the connectivity rule already implies
that a complex is present --- as under the AND-logic of
Sec.~\ref{sec:constructions}\,A, where a hyperedge survives only if all its
members do --- a complex reached along a functioning inclusion is present by
construction, and the occupation probability must not be folded into the map.
It is then reinstated at the root alone,
\begin{equation}
  P^l\to1-\pi_l+\pi_l P^l,
  \label{eq:rootocc}
\end{equation}
which again makes $S_l$ the fraction of \emph{all} layer-$l$ complexes.

\subsection{Solving the map}

\begin{table}[t]
\caption{\label{tab:code}Where each quantity is computed. Methods are on
\code{chygraph.Chygraph}; the constructors of Sec.~\ref{sec:constructions} are
functions in \code{chygraph.applications}.}
\begin{ruledtabular}
\begin{tabular}{ll}
quantity & method \\
\colrule
\multicolumn{2}{l}{\emph{closed form in the model parameters}}\\
map $F$, Eqs.~\eqref{eq:Qm}--\eqref{eq:Qp} & \code{apply} \\
$P^l$, Eq.~\eqref{eq:P}          & \code{root} \\
Jacobian $J$                     & \code{jacobian} \\
tensor $\vecA$, Eq.~\eqref{eq:A} & \code{A} \\
$\theta$                         & \code{theta} \\
$\Lambda$ and $\lambda$          & \code{Lambda}, \code{perron\_root} \\
amplitude $B$, Eq.~\eqref{eq:B}  & \code{amplitude} \\
$B$ on $\Lambda=0$               & \code{amplitude\_at\_threshold} \\
curvature $C$                    & \code{curvature} \\
core of the map                  & \code{core} \\
\colrule
\multicolumn{2}{l}{\emph{evaluated at a point in parameter space}}\\
fixed point $Q$                  & \code{solve} \\
$S_l$                            & \code{fractions} \\
$S=S_0$                          & \code{node\_fraction} \\
$B$, independent route           & \code{amplitude\_numeric} \\
core reduction check             & \code{verify} \\
\end{tabular}
\end{ruledtabular}
\end{table}

The only input specific to a construction is the four tables of generating
functions $\Phi^l_k$, $\bar\Phi^l_k$, $G^l_k$, $\bar G^l_k$: the same data the
threshold calculation needs, supplied as functions rather than as their first
derivatives at $1$. Everything downstream is common to all constructions.
Equations~\eqref{eq:Qm}--\eqref{eq:Qp} are assembled on the same $2L^2$ index
set for every one of them, and no equation specific to an example is written at
any point; each construction of Sec.~\ref{sec:constructions} is a dozen lines of
generating function definitions and nothing else.

The map is assembled by \code{Chygraph.apply} and \code{Chygraph.root} from the
four tables, and solved by iteration in \code{Chygraph.solve}. $F$ carries
$[0,1]^{2L^2}$ into itself and has
non-negative Taylor coefficients, so it is monotone: started from $Q=0$ the
iterates increase, are bounded above by $1$, and converge to the smallest fixed
point. That is the physical solution --- the one giving the largest giant
component consistent with the equations --- and it is $Q=1$ throughout the
subcritical phase. Substituting it into Eq.~\eqref{eq:P} gives $S_l$, which is
\code{Chygraph.fractions}; the fraction of nodes $S=S_0$ is
\code{Chygraph.node\_fraction}. Table~\ref{tab:code} lists the method behind
every quantity reported below.

Three features of the procedure are worth recording. Convergence is geometric
with a rate that tends to one as $\Lambda\to0$ from either side, so a tight
tolerance is needed near the threshold. Many of the $2L^2$ unknowns are
determined but decoupled, corresponding to states that cannot be reached in a
given construction; they are carried along harmlessly and do not affect $S_l$,
which is why the generic system is larger than the two or four equations one
would write by hand for a particular model. And the fixed point has no closed
form in general: what the calculation delivers symbolically is the map, and from
it the threshold diagnostics $\theta$ and $\Lambda$ of Sec.~\ref{sec:threshold}
and the amplitude $B$ of Sec.~\ref{sec:amplitude}, while $S$ at a given point in
parameter space is evaluated numerically.

This distinction is what separates the curves shown later. The solid lines in
Figs.~\ref{fig:triangles}, \ref{fig:correlated} and \ref{fig:applications} are
the fixed point itself, computed from the generic equations at each parameter
value and valid at any distance from the threshold. The linear behaviour
$B\Lambda$ is only a tangent at $\Lambda=0$; it is drawn separately, as the
dashed lines of Fig.~\ref{fig:degeneracy} and as the limit approached in
Fig.~\ref{fig:amplitude}.

\subsection{Checks}

\begin{figure}[t]
\includegraphics[width=\columnwidth]{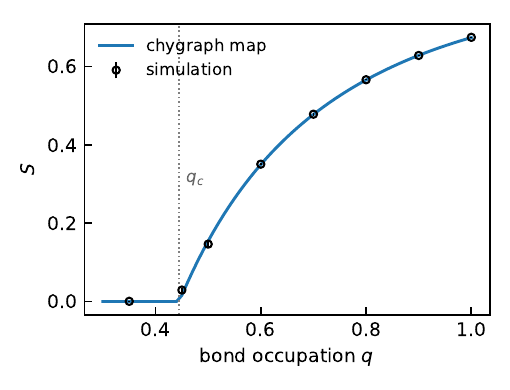}
\caption{\label{fig:triangles}Giant component fraction for bond percolation on a
graph with over-represented triangles, $\ave{k}_|=1$ and
$\ave{k}_\triangle=0.5$. The line is the non-linear fixed point of
Eqs.~\eqref{eq:Qm}--\eqref{eq:Qp}, valid at any distance from the threshold and
not the leading behaviour of Eq.~\eqref{eq:B}; symbols are Monte Carlo
simulation of configuration model graphs with $n=3\times10^5$ nodes, error bars
one standard deviation. The dotted line is the threshold $q_c=0.445042$ from
$\Lambda=0$.}
\end{figure}

The fixed point itself is checked against known closed forms and against direct
simulation. For graphs it reproduces $S=1-e^{-\ave{k}S}$ to machine precision.
Figure~\ref{fig:triangles} compares the map against Monte Carlo simulation of
the graph with over-represented triangles of Ref.~\cite{vazquez2024comnet}: a
configuration model in which each node carries a Poisson number $k_|$ of link
stubs and a Poisson number $k_\triangle$ of triangle corners, $n=3\times10^5$
nodes, over the whole range of the bond occupation probability; agreement is within one standard deviation of the simulation at
every point, including close to the threshold.

\section{The threshold as a linearisation}\label{sec:threshold}

\subsection{The published tensor}

From the four moment matrices of Sec.~\ref{sec:chygraphs} one assembles the
six-index tensor \cite{vazquez2024comnet}
\begin{align}
  (A_{--})^{ml}_{nk} &= \delta_{mn}\delta_{lk}-\ave{\kappa}_{nk}\delta_{nl},
  \nonumber\\
  (A_{-+})^{ml}_{nk} &= -\left[\ave{\bar s}_{nk}\delta_{mk}
                        +\ave{s}_{nk}(1-\delta_{mk})\right]\delta_{nl},
  \nonumber\\
  (A_{+-})^{ml}_{nk} &= -\left[\ave{\bar\kappa}_{nk}\delta_{mk}
                        +\ave{\kappa}_{nk}(1-\delta_{mk})\right]\delta_{nl},
  \nonumber\\
  (A_{++})^{ml}_{nk} &= \delta_{mn}\delta_{lk}-\ave{s}_{nk}\delta_{nl},
  \label{eq:A}
\end{align}
vectorised as $(\mathrm{vec}A)_{mL+l,\,nL+k}=A^{ml}_{nk}$ into the
$2L^2\times2L^2$ matrix $\vecA$. The order parameter of the transition is
$\Lambda=\max\mathrm{eig}(-\vecA)$, and $\theta=-\det(\vecA)$ is a
pseudo-order parameter with a simpler algebraic form. In the code the tensor is
\code{Chygraph.A} and the two diagnostics are \code{Chygraph.theta} and
\code{Chygraph.Lambda}, all three symbolic in the model parameters.

\subsection{$A$ is the Jacobian of the map}

Differentiating Eqs.~\eqref{eq:Qm}--\eqref{eq:Qp} at $Q=1$, where every
generating function equals $1$, leaves a single first derivative per term. Using
Eq.~\eqref{eq:firstderiv} and comparing with Eq.~\eqref{eq:A},
\begin{equation}
  \boxed{\;\vecA = I - J,\qquad
  J_{ab}=\left.\frac{\partial F_a}{\partial Q_b}\right|_{Q=1}\;}
  \label{eq:AisJ}
\end{equation}
entry by entry, in the same index layout. The four blocks of Eq.~\eqref{eq:A}
are the four ways of composing a step with the state it lands in, and the
Kronecker deltas $\delta_{mk}$ that select the excess averages are exactly the
$[\;\cdot\;]_{k=m}$ of Eqs.~\eqref{eq:Qm}--\eqref{eq:Qp}. Consequently
\begin{equation}
  \Lambda=\max\mathrm{eig}(-\vecA)=\lambda-1,
\end{equation}
where $\lambda$ is the Perron root of $J$: the threshold calculation is the
statement that the trivial fixed point $Q=1$ loses stability. Everything already
computed in Ref.~\cite{vazquez2024comnet} is recovered unchanged, and the order
parameter is obtained by solving the map instead of linearising it.

\subsection{Checks}

Equation~\eqref{eq:AisJ} has been verified entry by entry, symbolically, for
every mapping considered here, by comparing \code{Chygraph.A} against
\code{PercolationMatrix}, the implementation of the published calculation: the
Jacobian of
Eqs.~\eqref{eq:Qm}--\eqref{eq:Qp} reproduces the published $\vecA$ of
Eq.~\eqref{eq:A} exactly, and the Perron root reproduces the published
pseudo-order parameter through $\lambda^2=\theta+1$.

\section{What the first moments do not determine}

\begin{figure}[t]
\includegraphics[width=\columnwidth]{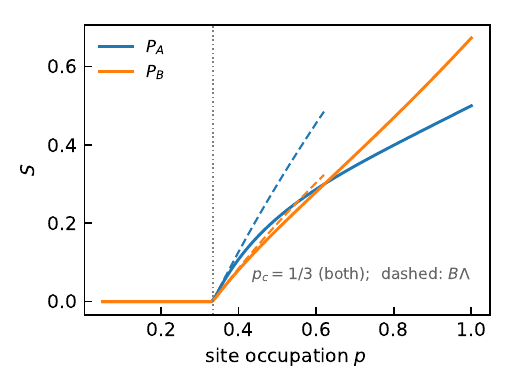}
\caption{\label{fig:degeneracy}First moments do not determine the order
parameter. Site percolation on two configuration model graphs with degree
distributions $P_A(k)=0.5\,\delta_{k0}+0.5\,\delta_{k4}$ and
$P_B(k)=0.2\,\delta_{k0}+0.5\,\delta_{k1}+0.3\,\delta_{k5}$, which share
$\ave{k}=2$ and $\ave{\kb}=3$ and therefore share the threshold $p_c=1/3$, but
differ in $S$ everywhere above it and in the critical amplitude. Dashed lines
are the leading behaviour $B\Lambda$ with $B$ from Eq.~\eqref{eq:Bgraph}.}
\end{figure}

Since $A$ depends on the generating functions only through
Eq.~\eqref{eq:firstderiv}, and moments do not determine a distribution, $S$
cannot be a function of $A$. The degeneracy is easy to exhibit. Take two graphs
with degree distributions
\begin{align}
  P_A(k)&=0.5\,\delta_{k0}+0.5\,\delta_{k4},\nonumber\\
  P_B(k)&=0.2\,\delta_{k0}+0.5\,\delta_{k1}+0.3\,\delta_{k5}.
\end{align}
Both have $\ave{k}=2$ and $\ave{\kb}=\ave{k(k-1)}/\ave{k}=3$, hence identical
$\theta=3pq-1$, identical $\Lambda$ and identical site percolation threshold
$p_c=1/3$ at $q=1$, yet at $p=q=1$ they give $S=0.500$ and $S=0.673$
(Fig.~\ref{fig:degeneracy}); both are supplied to \code{hypergraph\_giant} as
\code{finite\_pgf} tables and differ in no other input. They also differ in the
critical amplitude derived below, $B=4/3$ against $B=8/9$, which is visible as the different initial slopes
in Fig.~\ref{fig:degeneracy}.

The extra input is nonetheless cheap, because every example in
Ref.~\cite{vazquez2024comnet} was built from a distribution and then projected
onto a mean. The excess component sizes of a triangle under bond percolation,
for instance, were enumerated there in full,
\begin{align}
  \bar G_\triangle(y)=&\left(3q^2-2q^3\right)y^2+2q(1-q)^2\,y\nonumber\\
    &+(1-q)^3+q(1-q)^2,
  \label{eq:tripgf}
\end{align}
of which $\ave{\bar s}_{20}=\bar G_\triangle'(1)=2q(1+q-q^2)$ is the mean.
Un-projecting is a one-line change per model. For Poisson distributions the
generating function is fixed by its mean, $\Phi(x)=\bar\Phi(x)=
e^{\ave{\kappa}(x-1)}$, and the input collapses back to the matrices already in
use.

Reinstating the distribution also exposed an arithmetic slip in
Ref.~\cite{vazquez2024comnet}. Its Eq.~(30) for the mean intra-complex
component size of a triangle should carry $5/3$ rather than $3/2$ on the
single-link term,
\begin{equation}
  s_\triangle(q)=3\left(q^3+3q^2(1-q)\right)+5q(1-q)^2+(1-q)^3,
\end{equation}
equivalently $s_\triangle=1+\ave{\bar s}_{20}=1+2q(1+q-q^2)$, which is what
Eq.~\eqref{eq:tripgf} gives. The published threshold is unaffected, because
$\ave{s}_{20}$ never enters $\theta$: triangles are never reached from above.
The coefficient does enter the order parameter of the triangle layer, $S_2$.

\section{The critical amplitude}\label{sec:amplitude}

\begin{table*}[t]
\caption{\label{tab:amplitudes}Perron root $\lambda=1+\Lambda$ and critical
amplitude $B$ in $S=B\Lambda+O(\Lambda^2)$, on the critical manifold
$\lambda=1$. Sites are present with probability $p$, bonds or hyperedges with
probability $q$.}
\begin{ruledtabular}
\begin{tabular}{lll}
system & $\lambda$ & $B$ \\
\colrule
graph, arbitrary $P(k)$ &
  $\sqrt{pq\ave{\kb}}$ &
  $4p\ave{k}\ave{\kb}\big/\ave{\kb(\kb-1)}$ \\[2pt]
hypergraph, arbitrary $P(k)$, $P(c)$ &
  $\sqrt{pq\ave{\kb}\ave{\cb}}$ &
  $4p\ave{k}\ave{\kb}\ave{\cb}\big/
   \left[\ave{\cb}\ave{\kb(\kb-1)}+p\ave{\kb}^2\ave{\cb(\cb-1)}\right]$ \\[2pt]
Poisson graph &
  $\sqrt{pq\ave{k}}$ &
  $4p$ \\[2pt]
Poisson hypergraph &
  $\sqrt{pq\ave{k}\ave{c}}$ &
  $4p\big/\left(1+p\ave{c}\right)$ \\[2pt]
Poisson multiplex hypergraph, $L$ types &
  $\big(\sum_l q_l\ave{k}_l\ave{c}_l\big)^{1/2}$ &
  $4\big/\big(1+\sum_l q_l\ave{k}_l\ave{c}_l^2\big)$ \\[2pt]
Poisson graph with triangles, bond $q$ &
  $\big(q\ave{k}_|+2q(1+q-q^2)\ave{k}_\triangle\big)^{1/2}$ &
  $4\big/\left[1+2q^2(3-2q)\ave{k}_\triangle\right]$ \\
\end{tabular}
\end{ruledtabular}
\end{table*}

\subsection{Expansion of the map}

Write $Q=1-x$ and expand $F$ about the trivial fixed point,
\begin{equation}
  x_a=\sum_b J_{ab}x_b-\frac12\sum_{bc}M_{abc}x_bx_c+O(x^3),
  \label{eq:expansion}
\end{equation}
where $M_{abc}=\partial^2F_a/\partial Q_b\partial Q_c$ evaluated at $Q=1$.
Let $r$ and $\ell$ be the right and left Perron vectors of $J$, normalised by
$\ell\cdot r=1$. Just above the threshold $\lambda=1+\Lambda$ with
$\Lambda\to0^+$; setting $x=\epsilon r$ and projecting Eq.~\eqref{eq:expansion}
on $\ell$ removes the linear term and leaves
\begin{equation}
  \epsilon=\frac{2\Lambda}{C},\qquad C\equiv\ell\cdot M[r,r].
  \label{eq:epsilon}
\end{equation}
Expanding Eq.~\eqref{eq:P} to first order, $S_l=\nabla P^l\cdot x$, gives
\begin{equation}
  \boxed{\;S_l=B_l\Lambda+O(\Lambda^2),\qquad
  B_l=\frac{2\,\nabla P^l\cdot r}{C}\;}
  \label{eq:B}
\end{equation}
Only \emph{second} derivatives of the generating functions enter $B_l$, one
moment order beyond what $A$ requires. This is the middle row of
Table~\ref{tab:hierarchy}, and it is the reason the amplitude remains a symbolic
calculation while $S$ away from the threshold does not.

Two computational points make Eq.~\eqref{eq:B} practical. The rank-three tensor
$M$ never has to be formed: contracting it with $r$ twice is a second
directional derivative,
\begin{equation}
  M_a[r,r]=\left.\frac{d^2}{dt^2}F_a(1+tr)\right|_{t=0},
\end{equation}
a single-variable derivative of a product of generating functions, and likewise
$\nabla P^l\cdot r$ is a first directional derivative. Equation~\eqref{eq:B} is
\code{Chygraph.amplitude} and the contraction is done this way there. And the
$2L^2$ unknowns reduce: discarding iteratively every index whose row or column of $J$ vanishes
identically leaves the physically coupled \emph{core}, without changing the
non-zero spectrum. For a hypergraph the core is $2$ of $8$ indices, for a graph
with links and triangles $4$ of $18$, and in both cases the surviving indices
are precisely the unknowns one would have written down by hand. The reduction is
\code{Chygraph.core}, and \code{Chygraph.verify} checks it against the full
index set by solving the Perron problem there numerically.

\subsection{Closed forms}\label{sec:closedforms}

Table~\ref{tab:amplitudes} collects $B\equiv B_0$ on the critical manifold
$\lambda=1$ for the standard mappings, as returned by
\code{Chygraph.amplitude\_at\_threshold}. Only the first two derivatives of the
generating functions at $1$ enter, so an arbitrary distribution can be handed to
the calculation through \code{moment\_pgf}, a quadratic surrogate carrying its
first two factorial moments; the amplitudes below for arbitrary $P(k)$ and
$P(c)$ were obtained that way. Two results are worth isolating. For a
graph with an arbitrary degree distribution, sites present with probability $p$
and bonds with probability $q$,
\begin{equation}
  \lambda=\sqrt{pq\ave{\kb}},\qquad
  B=\frac{4p\,\ave{k}\ave{\kb}}{\ave{\kb(\kb-1)}},
  \label{eq:Bgraph}
\end{equation}
and for a hypergraph with arbitrary degree and cardinality distributions,
\begin{equation}
  \lambda=\sqrt{pq\ave{\kb}\ave{\cb}},\quad
  B=\frac{4p\,\ave{k}\ave{\kb}\ave{\cb}}
         {\ave{\cb}\ave{\kb(\kb-1)}+p\ave{\kb}^2\ave{\cb(\cb-1)}}.
  \label{eq:Bhyper}
\end{equation}
Equation~\eqref{eq:Bgraph} reproduces the two distributions of
Fig.~\ref{fig:degeneracy}: both have $\ave{k}=2$, $\ave{\kb}=3$ and $p_c=1/3$,
but $\ave{\kb(\kb-1)}=6$ and $9$ respectively, giving $B=4/3$ and $B=8/9$.
Setting the degree distribution Poisson in Eq.~\eqref{eq:Bgraph} gives $B=4p$,
and at $p=q=1$ the Erd\H{o}s--R\'enyi result $S\simeq4\Lambda=2(\ave{k}-1)$.

For Poisson multiplex hypergraphs the threshold is additive over hyperedge
types, $\theta=\sum_l q_l\ave{k}_l\ave{c}_l-1$
\cite{sun2021,vazquez2024comnet}. The amplitude inherits that structure,
\begin{equation}
  B=\frac{4}{1+\sum_l q_l\ave{k}_l\ave{c}_l^2},
  \label{eq:Bmultiplex}
\end{equation}
verified symbolically with \code{multiplex\_hypergraph\_giant} for $L=1,2,3$
hyperedge types: each layer contributes to
the denominator its own contribution to $\theta$, weighted by one further factor
of its cardinality.

For bond percolation on a graph with over-represented triangles, mapped to a
multiplex chygraph with a link layer and a triangle layer and Poisson
participation degrees, the intra-complex generating functions are the Bernoulli
$\bar G_|(y)=1-q+qy$ and Eq.~\eqref{eq:tripgf}. Then
\begin{equation}
  B=\frac{4}{1+2q^2(3-2q)\ave{k}_\triangle},
  \label{eq:Btriangle}
\end{equation}
which does not contain $\ave{k}_|$: the link layer contributes no curvature,
because a link's excess component size is Bernoulli and its generating function
is affine, so $\bar G_|''=0$. The link layer moves the threshold but not the
amplitude. This is a direct consequence of the fine-grained structure that
chygraphs make explicit, and has no counterpart at the level of $\theta$, where
both layers appear symmetrically.

\subsection{When the expansion fails}

The denominator $C=\ell\cdot M[r,r]$ of Eq.~\eqref{eq:B} is a sum of
non-negative terms, since $\ell$, $r$ and all second derivatives of generating
functions at $1$ are non-negative. When $C$ is finite and positive the
transition is continuous with $S\propto\Lambda$, that is with order parameter
exponent $\beta=1$, since $\Lambda$ vanishes linearly in the control parameter
at a generic threshold. Both qualifications matter.

Finiteness of $C$ is a condition on \emph{second} factorial moments of the
excess distributions, one order beyond what $\theta$ needs. For a graph
$C\propto\ave{\kb(\kb-1)}=\ave{k(k-1)(k-2)}/\ave{k}$, which diverges for a
scale-free degree distribution with exponent $\gamma\le4$. There
Eq.~\eqref{eq:B} gives $B\to0$: the linear branch has vanishing amplitude and
$\beta\ne1$, recovering the familiar $\beta=1/(\gamma-3)$ of the
heterogeneous mean field for $3<\gamma<4$. The expansion therefore signals its
own breakdown, but the closed forms of Table~\ref{tab:amplitudes} are
statements about ensembles with finite second moments and must not be read
outside that range.

When $C=0$ the map is affine along the critical direction and
Eq.~\eqref{eq:epsilon} has no linear branch; the elementary example is a
two-regular graph, where $\bar\Phi(x)=x$ and there is no transition of this
type at all. $C$ is therefore a symbolic continuity criterion, computable
alongside $\theta$ and $\Lambda$; it is \code{Chygraph.curvature}, with
\code{Chygraph.is\_continuous} reporting whether it vanishes identically. Its non-negativity also settles, for anything
that maps to a chygraph, the recent claim of Keating and H\'ebert-Dufresne
\cite{keating2026} that group structure on its own produces only continuous
transitions: a complex contributes to $C$ through second derivatives of its
generating functions, which cannot be negative.

A caveat delimits the scope. Equations~\eqref{eq:Qm}--\eqref{eq:Qp} are the
extinction conditions of a multi-type branching process, and for such maps the
loss of stability of $Q=1$ locates the transition. Interdependent networks
\cite{buldyrev2010,radicchi2015} are not of this class: their giant component
appears through a saddle-node bifurcation while $Q=1$ is still stable, so
$\Lambda=0$ does not locate the transition, and the physical fixed point is
reached by a cascade from above rather than by monotone iteration from $Q=0$.
Mapping those constructions to chygraphs, and identifying what replaces
$\Lambda$ there, is left for future work.

\subsection{Checks}

\begin{figure}[t]
\includegraphics[width=\columnwidth]{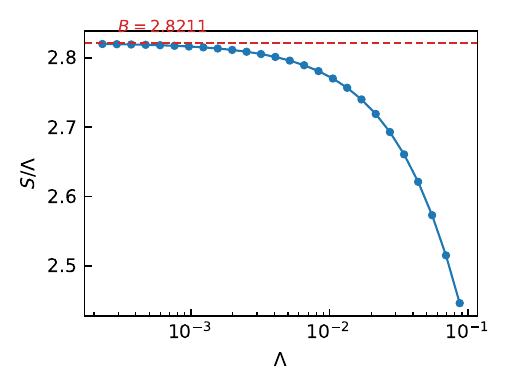}
\caption{\label{fig:amplitude}Approach to the critical amplitude for the graph
with over-represented triangles of Fig.~\ref{fig:triangles}. $S/\Lambda$ from
the solved map against $\Lambda$, as the threshold is approached from above,
converging to the closed form
$B=4/[1+2q_c^2(3-2q_c)\ave{k}_\triangle]=2.8211$ of Eq.~\eqref{eq:Btriangle}
(dashed line).}
\end{figure}

The closed-form amplitudes are checked against the numerically solved map.
Figure~\ref{fig:amplitude} shows $S/\Lambda$ approaching $B=2.8211$ from
Eq.~\eqref{eq:Btriangle} as the threshold $q_c=0.445042$ is approached from
above, for the same graph with over-represented triangles as
Fig.~\ref{fig:triangles}. They are further cross-checked against
\code{Chygraph.amplitude\_numeric}, an independent evaluation of
Eq.~\eqref{eq:B} by numerical eigenvectors on the full, unreduced index set,
confirming that the core reduction is exact.

\section{Dependent layers}\label{sec:joint}

Section~\ref{sec:map} took the generating functions to factorise over target layers,
$\Phi^l(x_0,\dots,x_{L-1})=\prod_k\Phi^l_k(x_k)$. That covers every mapping in
Ref.~\cite{vazquez2024comnet}, but not constructions in which a complex's
participation in different layers is correlated: a node whose number of
triangles is tied to its number of links, a hyperedge whose cardinalities in two
vertex layers move together. Joint degree distributions of this kind are exactly
what the motif formalism of Mann \textit{et al.} \cite{mann2021} works with.

Dropping the factorisation changes the status of the excess generating
functions. They are no longer independent inputs but derivatives of the joint
one,
\begin{equation}
  \bar\Phi^{l,(m)}(x)=\frac{\partial\Phi^l/\partial x_m}
                           {\left.\partial\Phi^l/\partial x_m\right|_{x=1}},
  \label{eq:excessderived}
\end{equation}
and likewise $\bar G^{l,(m)}$. The map becomes
\begin{align}
  Q^{ml}_- &= \Phi^l\!\left(Q^{l\cdot}_-\right)\,
              \bar G^{l,(m)}\!\left(Q^{l\cdot}_+\right),\nonumber\\
  Q^{ml}_+ &= \bar\Phi^{l,(m)}\!\left(Q^{l\cdot}_-\right)\,
              G^l\!\left(Q^{l\cdot}_+\right),
  \label{eq:jointmap}
\end{align}
with $P^l=\Phi^l(Q^{l\cdot}_-)G^l(Q^{l\cdot}_+)$, reducing term by term to
Eqs.~\eqref{eq:Qm}--\eqref{eq:Qp} when the generating functions factorise. This
is \code{JointChygraph}, which takes the joint functions in place of the four
factorised tables and derives the excess ones; it shares the whole of
Table~\ref{tab:code} with the factorised class, differing only in how the
chygraph is specified.

\subsection{The generalised threshold tensor}

Differentiating Eq.~\eqref{eq:jointmap} at $Q=1$ and using
Eq.~\eqref{eq:excessderived}, the second derivative of the joint generating
function appears where the published tensor carries a first moment. Define the
\emph{inclusion-biased} moment matrices
\begin{align}
  \ave{\bar\kappa^{(m)}}_{lk}
    &=\frac{\ave{\kappa_{lm}\left(\kappa_{lk}-\delta_{mk}\right)}}
           {\ave{\kappa_{lm}}}
     =\frac{1}{\ave{\kappa}_{lm}}
      \left.\frac{\partial^2\Phi^l}{\partial x_m\partial x_k}\right|_{x=1},
      \nonumber\\
  \ave{\bar s^{(m)}}_{lk}
    &=\frac{\ave{s_{lm}\left(s_{lk}-\delta_{mk}\right)}}{\ave{s_{lm}}},
  \label{eq:biased}
\end{align}
the expected chy-degree to layer $k$, and intra-complex component size in layer
$k$, of a complex reached through an inclusion into layer $m$. Then the Jacobian
of Eq.~\eqref{eq:jointmap} is Eq.~\eqref{eq:A} with
\begin{align}
  (A_{-+})^{ml}_{nk} &= -\ave{\bar s^{(m)}}_{nk}\,\delta_{nl},\nonumber\\
  (A_{+-})^{ml}_{nk} &= -\ave{\bar\kappa^{(m)}}_{nk}\,\delta_{nl},
  \label{eq:Ajoint}
\end{align}
one expression per block in place of the $\delta_{mk}$ / $(1-\delta_{mk})$ split.
At $k=m$, Eq.~\eqref{eq:biased} is the excess average and
Eq.~\eqref{eq:Ajoint} agrees with Eq.~\eqref{eq:A} identically. At $k\ne m$ it
agrees only when $\ave{\kappa_m\kappa_k}=\ave{\kappa_m}\ave{\kappa_k}$. The
published tensor is therefore the independent-layer special case; when the
layers are correlated its off-diagonal slots must carry the joint second moment,
and both the threshold and the order parameter change.

One caveat on Eq.~\eqref{eq:excessderived}. It is the configuration model
size-biasing relation, correct whenever a complex is reached by sampling one of
its inclusions uniformly, which is always the case for the chy-degrees
$\Phi^l$. It is \emph{not} correct when the fine-grained structure inside a
complex makes the entry vertex matter, that is when $G^l$ generates a component
size rather than a cardinality. For a triangle under bond percolation
$\bar G_\triangle$ is Eq.~\eqref{eq:tripgf} and not $G'_\triangle/G'_\triangle(1)$,
and it must be supplied rather than derived. This is precisely the distinction
between the chy-degree structure and the intra-complex structure that chygraphs
separate by construction.

\begin{table*}[t]
\caption{\label{tab:mappings}Chygraph mappings. Layer 0 always holds the
elementary units as atoms. $\bar G_c$ is the excess cardinality generating
function, $\bar\Phi_k$ the excess chy-degree generating function, and
$\bar G_{K_n}$ the excess of the bond percolation cluster size in a complete
graph $K_n$. Where a cell holds two entries they are the generating functions
of the two complex layers, in the order the layers are listed. The constructor
that supplies each mapping is shown in Fig.~\ref{fig:code}.}
\begin{ruledtabular}
\begin{tabular}{llll}
construction & layers & $\Phi^0$ & $\bar G^l_0$ \\
\colrule
OR / factor graph percolation &
  0 nodes, 1 hyperedges &
  $\Phi_k(x_1)$ &
  $\bar G_c(1-p+py)$ \\[2pt]
AND / hypergraph percolation &
  0 nodes, 1 hyperedges &
  $\Phi_k(x_1)$ &
  $1-\bar G_c(p)+\bar G_c(py)$ \\[2pt]
hyperdegree--cardinality correlation &
  0 nodes, $1..L$ hyperedges by cardinality &
  $\Phi(x_1,\dots,x_L)$ joint &
  $(1-p+py)^{c_l-1}$ \\[2pt]
two levels of mixing &
  0 people, 1 households, 2 global contacts &
  $x_1\,\Phi_k(x_2)$ &
  \begin{tabular}[t]{@{}l@{}}$\bar G_{K_n}(y)$\\$1-T+Ty$\end{tabular} \\[2pt]
network of cliques / bipartite projection &
  0 nodes, 1 cliques &
  $\Phi_k(x_1)$ &
  $\bar G_{K_n}(y)$ \\[2pt]
triadic closure ($R=2$) &
  0 occupied, 1 vacant, 2 edges &
  $g_0(x_2)$ &
  \begin{tabular}[t]{@{}l@{}}$p\,y_0+(1-p)y_1$\\$p\,y_0+1-p$\end{tabular} \\
\end{tabular}
\end{ruledtabular}
\end{table*}

\subsection{Identical marginals, different transitions}

\begin{figure}[t]
\includegraphics[width=\columnwidth]{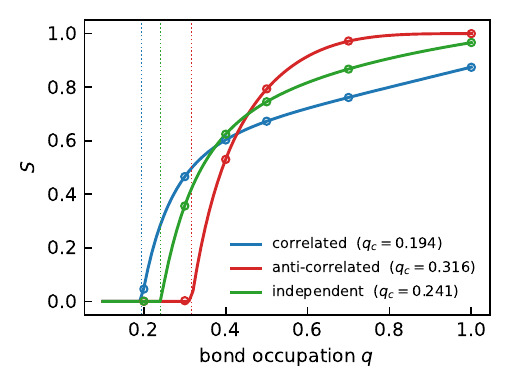}
\caption{\label{fig:correlated}Correlated layers. Bond percolation on a graph
with links and triangles, for three joint distributions of the number of links
$k_|$ and triangles $k_\triangle$ a node participates in: $(k_|,k_\triangle)$
equal to $(1,0)$ or $(3,2)$ with probability $1/2$ each (correlated), $(1,2)$ or
$(3,0)$ with probability $1/2$ each (anti-correlated), and the product of the
two marginals $k_|\in\{1,3\}$, $k_\triangle\in\{0,2\}$ (independent). All
three share the marginals, hence $\ave{k}_|=2$, $\ave{\kb}_|=3/2$,
$\ave{k}_\triangle=1$, $\ave{\kb}_\triangle=1$, and every entry of the
published threshold tensor. Lines are Eq.~\eqref{eq:jointmap}, symbols are Monte
Carlo simulation with $n=3\times10^5$ nodes, dotted lines the thresholds from
Eqs.~\eqref{eq:biased}--\eqref{eq:Ajoint}. The first moments predict the
independent curve throughout.}
\end{figure}

The effect is not a small correction. Take bond percolation on a graph with
links and triangles, and three joint distributions of the number of links $k_|$
and triangles $k_\triangle$ a node participates in, all with the same marginals
$k_|\in\{1,3\}$ and $k_\triangle\in\{0,2\}$ with probability $1/2$ each, hence
the same $\ave{k}_|=2$, $\ave{\kb}_|=3/2$, $\ave{k}_\triangle=1$,
$\ave{\kb}_\triangle=1$, and the same published $\theta$:
\begin{align}
  \text{correlated:}&\quad (k_|,k_\triangle)\in\{(1,0),(3,2)\},\nonumber\\
  \text{anti-correlated:}&\quad (k_|,k_\triangle)\in\{(1,2),(3,0)\},\nonumber\\
  \text{independent:}&\quad \text{the product of the marginals.}
\end{align}
Equation~\eqref{eq:biased} separates them through
$\ave{\bar\kappa^{(1)}}_{0\triangle}=\ave{k_|k_\triangle}/\ave{k_|}$, which is
$3/2$, $1/2$ and $1$ respectively, the last coinciding with
$\ave{\kappa}_{0\triangle}=1$ as it must. The three bond percolation thresholds
are, in the same order,
\begin{equation}
  q_c=0.1939,\qquad 0.3161,\qquad 0.2409,
  \label{eq:qc3}
\end{equation}
and the order parameters differ everywhere above them,
Fig.~\ref{fig:correlated}. All three agree with simulation; the marginals alone
predict the middle curve for all three.

The mechanism is intuitive. Positive correlation concentrates both kinds of
connectivity on the same nodes, which lowers the threshold --- those nodes form
a well-connected core --- but caps the reachable fraction, because the
complementary nodes are poorly connected in both layers. Anti-correlation
spreads connectivity out, raising the threshold but eventually pulling every
node in, $S\to1$ at $q=1$. Neither is visible in the first moments.

The construction is checked in both directions. With factorised inputs its
Jacobian reproduces the published $\vecA$ entry by entry and its fixed point
reproduces the factorised map of Sec.~\ref{sec:map} to machine precision, for both the
hypergraph and the graph with triangles. With correlated inputs it is compared
against configuration model simulation with the prescribed joint degree pairs,
over all three distributions and the whole range of $q$: agreement is within one
standard deviation in every case, while the marginals-only prediction is outside
it in six of nine.

\subsection{Uniqueness of the giant component}

The fixed point returns the fraction of complexes in an \emph{infinite}
component, and percolation theory identifies that with the giant component under
the assumption that the infinite component is unique. Correlated layers can
break that assumption. If $\Phi^0(x)=\tfrac12x_|^2+\tfrac12x_\triangle^2$, no
node participates in both layers, the chygraph splits into two disjoint pieces,
and the map returns the sum of their infinite components rather than the larger
one: at $q=1$ it gives $S=1$ where simulation gives $1/2$. The decomposition is
easy to check for and does not arise when some layer is on its own supercritical
and touches every complex, which is the case for all three distributions in
Fig.~\ref{fig:correlated}.

\section{Constructions from the literature}\label{sec:constructions}

The value of the map is that a published percolation or spreading problem is
solved by writing down its chygraph mapping and substituting. This section does
that for four families, each of which has been treated separately with a
purpose-built calculation. The mappings are collected in
Table~\ref{tab:mappings} and the resulting order parameters in
Fig.~\ref{fig:applications}. In every case the only work is identifying
$\Phi^l$ and $G^l$; the threshold, the order parameter and the amplitude then
follow from Secs.~\ref{sec:map}--\ref{sec:joint} unchanged. Figure~\ref{fig:code}
shows the inputs as they are actually supplied: one constructor call per
construction, returning a \code{Chygraph} that the methods of
Table~\ref{tab:code} then act on without further work.

\begin{figure*}[t]
\small
\begin{verbatim}
from chygraph import *

# AND- against OR-logic hypergraph percolation.  One chygraph, two intra-complex
# generating functions; the degree and cardinality PGFs are the optional arguments.
and_or_hypergraph('or')                        # Gbar = Gbar_c(1 - p + p y)
and_or_hypergraph('and')                       # Gbar = 1 - Gbar_c(p) + Gbar_c(p y)

# Hyperedges split into layers by cardinality class; the correlation between a node's
# hyperdegree and the cardinalities it joins is the joint chy-degree generating function.
Phi = two_class_joint_degree(1, Rational(3, 5), Rational(4, 5), w)
correlated_cardinality_hypergraph([2, 5], Phi)  # cardinality classes c = 2 and c = 5

# Households as complete graphs, global contacts as two-member complexes.  The argument
# is the household size distribution; Gbar is built from K_n internally.
household_epidemic({3: 1})                     # all households of size three
household_epidemic({2: Rational(1,2), 4: Rational(1,2)})
clique_network({3: 1}, p_bond=q)               # same complexes, many memberships per node

# Static triadic closure: the chygraph lives on the backbone, not on the clustered graph.
stc_percolation()                              # Poisson backbone; pass backbone_degree otherwise

# Every one then answers the same three questions through the same methods.
M = household_epidemic({3: 1})
M.theta()                     # T*k*(2*p_H - 2*p_H**3 + 2*p_H**2 + 1) - 1,  i.e. R* - 1
M.Lambda()                    # order parameter of the threshold, symbolic
M.amplitude()                 # B in S = B*Lambda + O(Lambda**2), symbolic
M.node_fraction({'p_H': 0.5, 'T': 0.3, 'k': 2.0})       # 0.41961, the final size
\end{verbatim}
\caption{\label{fig:code}How each construction is fed to the code. A constructor
returns a \code{Chygraph}, whose only construction-specific content is the four
tables of generating functions it was built from; the blocks follow the
subsections of Sec.~\ref{sec:constructions} in order; every method in
Table~\ref{tab:code} is then available on it without further work. The
constructors themselves are ten to twenty lines of generating function
definitions, with the mapping recorded in the docstring.}
\end{figure*}

\subsection{AND-logic against OR-logic}

Bianconi and Dorogovtsev \cite{bianconi2024} distinguish \emph{factor graph}
percolation, in which a higher order interaction keeps connecting whichever of
its members survive, from \emph{hypergraph} percolation, in which the
interaction fails outright if any one of its members is removed --- the right
rule for a chemical reaction, a catalytic step or a supply chain. Ha, Neri and
Annibale \cite{ha2025} call the same distinction OR- and AND-logic.

Both are the same chygraph: nodes as atoms in layer 0, hyperedges as complexes
in layer 1. They differ in one generating function. Under OR each of the
$c-1$ other members is reached when present; under AND all of them must be
present and then all are reached:
\begin{equation}
  \bar G^1_0(y)=
  \begin{cases}
    \bar G_c(1-p+py), & \text{OR},\\[2pt]
    1-\bar G_c(p)+\bar G_c(py), & \text{AND}.
  \end{cases}
  \label{eq:andor}
\end{equation}
Neither $\Phi^0_1$ nor $\bar\Phi^0_1$ is thinned: under either rule a node
reached along a functioning hyperedge is present by construction. Node
occupation reappears only at the root, through Eq.~\eqref{eq:rootocc}, so that
$S$ is the fraction of all nodes and $S\le p$.

Spelled out as input, the four tables are: $\Phi^0_1$ the hyperdegree generating
function and $\bar\Phi^0_1$ its excess, neither thinned; $G^1_0$ the cardinality
generating function; and $\bar G^1_0$ from Eq.~\eqref{eq:andor}, the one slot in
which the two rules differ. Every other slot is empty, since hyperedges are
included in nothing and nodes contain nothing. This is
\code{and\_or\_hypergraph}, which takes the logic as its first argument and the
degree and cardinality generating functions as optional ones, defaulting to
Poisson. Nothing downstream is aware of which rule was chosen: the $2L^2=8$
equations, the Jacobian, the determinant and the fixed point are the generic
ones of Secs.~\ref{sec:map} and~\ref{sec:threshold}.

Differentiating Eq.~\eqref{eq:andor} at $y=1$ gives the whole difference at the
level of the threshold tensor:
\begin{equation}
  \ave{\bar s}_{10}=p\,\bar G_c'(1)\ \ \text{(OR)},\qquad
  \ave{\bar s}_{10}=p\,\bar G_c'(p)\ \ \text{(AND)},
  \label{eq:andors}
\end{equation}
the same expression evaluated at $1$ and at $p$. Since $\bar G_c'$ is a power
series with non-negative coefficients it is non-decreasing on $[0,1]$, so
$\bar G_c'(p)\le\bar G_c'(1)$ and
\begin{equation}
  \theta_{\rm AND}\le\theta_{\rm OR},\qquad p_c^{\rm AND}\ge p_c^{\rm OR},
\end{equation}
with equality if and only if $p=1$ or $\bar G_c'$ is constant, that is
$c\equiv2$: for an ordinary graph the two rules are the same statement. This is
a one-line proof of the inequality reported in Ref.~\cite{bianconi2024}. For
Poisson degrees and cardinalities the two thresholds are
\begin{equation}
  \theta_{\rm OR}=p\ave{k}\ave{c}-1,\qquad
  \theta_{\rm AND}=p\ave{k}\ave{c}\,e^{\ave{c}(p-1)}-1,
\end{equation}
and the order parameters are shown in Fig.~\ref{fig:applications}(a): at
$\ave{k}=2$,
$\ave{c}=3$ the thresholds are $p_c=1/6$ and $p_c=0.5827$, and the two curves
meet at $p=1$ as they must.

\subsection{Hyperdegree--cardinality correlation}

Whether nodes of high hyperdegree preferentially join hyperedges of large
cardinality is a measurable property of real hypergraphs, and one that is known
to matter for connectivity \cite{ha2025,valdez2026}; the same effect arises in
bipartite projections through clique size fluctuations \cite{fujiki2024}. It is
a correlation between layers in the sense of Sec.~\ref{sec:joint}. Split the hyperedges into
layers by cardinality class $c_l$; then the joint chy-degree generating function
$\Phi^0(x_1,\dots,x_L)$ of the vector counting how many hyperedges of each class
contain a node \emph{is} the correlation, and it enters through the
inclusion-biased moments of Eq.~\eqref{eq:biased}.

To isolate it, take two classes $c=2$ and $c=5$ and the one-parameter family in
which a node is one of two equally likely types, and the types pair the high
mean in one class with the high or the low mean in the other with probability
$w$ and $1-w$. Both marginal hyperdegree distributions, and hence every entry
of the published threshold tensor, are the same for every $w$, while
$\mathrm{Cov}(\kappa_1,\kappa_2)$ runs linearly from maximally negative at $w=0$
through independent at $w=1/2$ to maximally positive at $w=1$.

The input is therefore a list of cardinalities, one per hyperedge layer, and a
single joint generating function of the chy-degree vector; the excess functions
are not supplied but derived from it by Eq.~\eqref{eq:excessderived}, and the
intra-complex functions $\bar G^l_0(y)=(1-p+py)^{c_l-1}$ follow from the
cardinality of each class. In the code the family is
\code{two\_class\_joint\_degree} and the chygraph is
\code{correlated\_cardinality\_hypergraph}; the inclusion-biased moments that
separate its members are read off with \code{JointChygraph.kappa\_bar}, and
\code{JointChygraph.layers\_independent} reports whether the published tensor
would have sufficed. The site percolation thresholds are
\begin{equation}
  p_c=0.2460,\qquad 0.2120,\qquad 0.1793
\end{equation}
at $w=0$, $1/2$, $1$: positive correlation concentrates connectivity and lowers
the threshold. The order parameter is not ordered the same way at all $p$. The
three curves cross, Fig.~\ref{fig:applications}(b): at $p=1/2$ the independent
member has the \emph{largest} giant component, $S=0.2423$ against $0.2204$
and $0.2369$, while at $p=1$ the ordering is monotonic in $w$, $S=0.7427$, $0.6698$,
$0.5905$. No single statement that correlation helps or hurts survives; the
threshold and the order parameter answer differently, which is an argument for
computing both.

\subsection{Epidemics with two levels of mixing}

An SIR epidemic with a constant infectious period is bond percolation
\cite{newman2002}, so a population in which individuals mix both within
households and along a global contact network --- the two levels of mixing model
of Ball, Mollison and Scalia-Tomba \cite{ball1997,ball2010} --- is a percolation
problem. It is a three-layer chygraph: individuals as atoms, households as
complexes whose intra-complex graph is complete, and global contacts as
complexes holding two individuals.

Two features of the mapping carry the content. First,
\begin{equation}
  \Phi^0_1(x)=x,\qquad\text{hence}\qquad \bar\Phi^0_1=1,
\end{equation}
because an individual belongs to exactly one household: reached from their
household they have no second household to pass through, so
$\ave{\bar\kappa}_{01}=0$ and the household layer is a sink rather than a route.
Households alone never percolate, $\theta=-1$ for every $p_H$, and the global
layer is what connects them. Second, a Reed--Frost epidemic in a closed group of
$n$ individuals with per-pair transmission probability $p_H$ is bond percolation
on $K_n$, so the within-household final size distribution --- the classical
object of Ref.~\cite{ball1986} --- is exactly $\bar G^1_0$, obtained from
\begin{equation}
  P_n(j)=\binom{n-1}{j-1}\,C_j(p_H)\,(1-p_H)^{j(n-j)},
  \label{eq:Kn}
\end{equation}
with $C_j$ the probability that $G(j,p_H)$ is connected, mixed over the
size-biased household size distribution because a household is reached through
one of its members. For $n=3$, Eq.~\eqref{eq:Kn} reproduces the triangle
enumeration of Ref.~\cite{vazquez2024comnet}, Fig.~3, term by term: the
household model and the over-represented-triangle model share their
intra-complex generating function.

The user supplies only the household size distribution: \code{household\_epidemic}
takes a dictionary over sizes, together with the within-household and global
transmission probabilities and the global degree generating function. From the
size distribution it builds $\bar G^1_0$ through
\code{clique\_excess\_pgf}, which evaluates Eq.~\eqref{eq:Kn} for each size and
mixes them size-biased, and fills the remaining slots with $\Phi^0_1(x)=x$, the
global degree functions in $\Phi^0_2$, and the Bernoulli $\bar G^2_0=1-T+Ty$. The
epidemiology enters through one table entry and nothing else.

Substituting into the tensor gives
\begin{equation}
  \theta+1=T\left[\ave{\kb}+\mu_H\ave{k}\right]\equiv R^*,
  \label{eq:Rstar}
\end{equation}
with $\mu_H=\bar G^{1\prime}_0(1)$ the mean number of additional household
members infected, verified symbolically for several household size
distributions. Equation~\eqref{eq:Rstar} is the household reproduction number of
Ref.~\cite{ball1997}, recovered here from the generic chygraph tensor with no
epidemiological input: a household reached by one global infection contains
$1+\mu_H$ cases, the index case makes $T\ave{\kb}$ further global infections and
each secondary case $T\ave{k}$. What the chygraph calculation adds is the final
size of a major outbreak, $S$, Fig.~\ref{fig:applications}(c), and its critical
amplitude, neither of which follows from the branching process argument that
gives $R^*$.

The same construction with $\Phi^0_1$ a genuine degree distribution rather than
the constant $1$ is a network of cliques, which covers SIR with clique-type
dependent transmission and, at $p_H=1$, the connected components of bipartite
projections \cite{fujiki2024}. With all cliques of size two it reduces to
ordinary bond percolation and with size three to the triangle layer of
Sec.~\ref{sec:closedforms}, two independent code paths that agree.

\subsection{Clustered graphs with overlapping loops}\label{sec:stc}

The mappings above all place the loops \emph{inside} complexes, which is where
the formalism can absorb them. It is worth showing what happens when they are
not, because a recent result makes the point sharply. Cirigliano
\cite{cirigliano2025} solves site percolation on Static Triadic Closure (STC)
graphs, built by closing every triad of a treelike backbone $G_0$, and finds
that heterogeneous mean field --- the treelike rewiring that preserves only the
degree distribution of the clustered graph $G_1$ --- gets both the threshold and
the critical exponents wrong.

The obvious chygraph for $G_1$ fails too. Closing all triads makes every closed
neighbourhood a clique, so $G_1$ is a hypergraph with nodes as atoms and
neighbourhoods as complexes. But every backbone edge $u\sim v$ then creates the
four-cycle $u-C(v)-v-C(u)-u$ in the node/complex incidence graph, and chygraphs
are exact only when that incidence structure is locally treelike. For a Poisson
backbone with $\ave{k}=3$ this mapping gives $p_c=0.0711$ against the exact
$0.0893$, no improvement on heterogeneous mean field's $0.0635$.

The formalism does cover the problem, but on the backbone rather than on $G_1$.
Site percolation on $G_1$ is equivalent to extended-range percolation with
$R=2$ on $G_0$, in which two occupied nodes are connected when at most one
unoccupied node lies between them. Promoting the occupation state to a layer
gives a three-layer chygraph: occupied nodes, unoccupied nodes and backbone
edges, with $\Phi^0=\Phi^1=g_0(x_2)$ and
\begin{align}
  \bar G^{2,(0)}(y)&=p\,y_0+(1-p)\,y_1,\nonumber\\
  \bar G^{2,(1)}(y)&=p\,y_0+(1-p).
  \label{eq:stc}
\end{align}
The entire range-2 rule is the difference between these two lines: entering an
edge from an unoccupied node, a second unoccupied node in a row kills the path.
That the intra-complex generating function may depend on the layer the complex
was entered from is the same freedom Sec.~\ref{sec:joint} required for motifs,
and it is why this construction is built with \code{JointChygraph} rather than
the factorised class: \code{stc\_percolation} passes the two functions of
Eq.~\eqref{eq:stc} as the $\bar G^{2,(0)}$ and $\bar G^{2,(1)}$ entries of the
\code{Gbar} table, supplies $g_0$ for both node layers, and sets the root
occupation of layer 0 to $p$. The backbone degree generating function is the
only physical input.

The tensor then returns
\begin{equation}
  \theta=-b^2p^2+b(1+b)p-1,\quad
  p_c=\frac{(1+b)-\sqrt{(1+b)^2-4}}{2b},
\end{equation}
with $b=\ave{k(k-1)}/\ave{k}$, which is Eq.~(12) of Ref.~\cite{cirigliano2025},
and an order parameter agreeing with that reference to machine precision at
every $p$. The exponents follow: $C$ is finite when the \emph{backbone}
third moment converges, giving $\beta=1$, and diverges otherwise, giving
$\beta=1/(\gamma_d-3)$. Written in terms of the STC degree exponent
$\tilde\gamma_d=\gamma_d-1$ these are $\beta=1$ for $\tilde\gamma_d>3$ and
$\beta=1/(\tilde\gamma_d-2)$ for $2<\tilde\gamma_d<3$, which is exactly the
table of exact exponents in Ref.~\cite{cirigliano2025}.

The moral is not that mean field always holds. It is that the tree-based
calculation was being applied to the wrong object: what fails is heterogeneous
mean field on $G_1$, not the generating function method. Mapping to a chygraph
moves the boundary of what counts as local from ``no loops'' to ``no loops
between complexes'', and finding a representation that respects the second
condition is the work --- here it required leaving $G_1$ altogether.

\subsection{Checks}

\begin{figure*}[t]
\includegraphics[width=\textwidth]{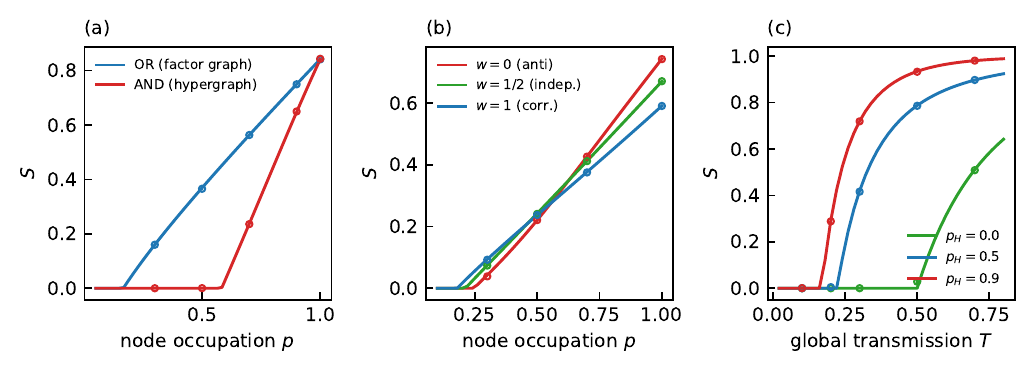}
\caption{\label{fig:applications}Three constructions mapped to chygraphs. Lines
are the fixed point of the chygraph map, Eqs.~\eqref{eq:Qm}--\eqref{eq:Qp} in
(a) and (c) and Eq.~\eqref{eq:jointmap} in (b); symbols are Monte Carlo
simulation ($n=1.5\times10^5$). (a) Site percolation on a
Poisson hypergraph, $\ave{k}=2$, $\ave{c}=3$, under OR-logic and AND-logic: the
same chygraph with the two generating functions of Eq.~\eqref{eq:andor}.
(b) Site percolation with hyperdegree--cardinality correlation, for three members
of a family with identical marginals; the curves cross, so the ordering at the
threshold reverses at large $p$. (c) Final size of an SIR epidemic with
households of size three and a Poisson global network, $\ave{k}=2$, against the
global transmission probability.}
\end{figure*}

Each construction is checked against simulation of the corresponding random
ensemble, Fig.~\ref{fig:applications}, and against independent analytical
results where they exist: the factor graph threshold
$p\ave{\kb}\ave{\cb}=1$, the household reproduction number
Eq.~\eqref{eq:Rstar}, the two reductions of the clique network just quoted, the
identity of Eq.~\eqref{eq:Kn} at $n=3$ with the published triangle enumeration,
and the closed-form STC threshold. The STC order parameter is
additionally checked against an independent iteration of the equations of
Ref.~\cite{cirigliano2025}, and those against Monte Carlo simulation of STC
graphs grown from a Poisson configuration-model backbone with $n=1.2\times10^5$
nodes, agreeing within one standard deviation at every $p$.

\section{Conclusions}

The tensor $A$ that gives the percolation threshold of a chygraph is the
Jacobian of a non-linear self-consistency map at its trivial fixed point. The
threshold and the order parameter are therefore not two calculations but two
orders of one, on the same index set and in the same vectorised layout, and the
symbolic calculator that returns $\theta$ and $\Lambda$ extends to the giant
component fraction with no new formalism.

What changes is the input. The threshold is fixed by the first moments of the
chy-degrees and intra-complex component sizes; the order parameter away from the
threshold requires the generating functions themselves, and no function of the
four moment matrices can supply it. Between those two lies the critical
amplitude, Eq.~\eqref{eq:B}, which needs only second moments and remains a
closed-form symbolic calculation. It is compact enough to be read: $4p$ for a
Poisson graph, $4p/(1+p\ave{c})$ for a Poisson hypergraph, additive over layers
for Poisson multiplex hypergraphs, and independent of the link layer for graphs
with over-represented triangles.

That last case shows what the representation buys. The threshold treats links
and triangles symmetrically; the amplitude does not, because a link's
intra-complex component size is Bernoulli and contributes no curvature.
Separating the large scale connectivity from the connectivity within a complex
is exactly the decomposition chygraphs are built to express, and the amplitude
is where that separation first becomes visible in the order parameter.

Writing the map also exposes an assumption the tensor had hidden. Independence
of a complex's participation across layers is not part of the chygraph
definition, and dropping it replaces the unconditional first moments in two of
the four blocks by inclusion-biased second moments,
Eqs.~\eqref{eq:biased}--\eqref{eq:Ajoint}, in a single expression that contains
the published one. Identical marginals with different joint structure give
different thresholds and different order parameters, so for constructions where
layer participation is correlated --- motif percolation with joint degrees among
them --- the marginals are not sufficient data.

The constructions of Sec.~\ref{sec:constructions} show what this buys in
practice. AND- and
OR-logic hypergraph percolation are one chygraph with two intra-complex
generating functions, and comparing them at the level of $\ave{\bar s}_{10}$
proves in one line an inequality that was established by separate calculation.
The two levels of mixing model of mathematical epidemiology is a three-layer
chygraph whose intra-complex generating function is the classical
within-household final size distribution, and its household reproduction number
is the chygraph threshold; the final size, which the branching process argument
does not give, is the chygraph order parameter. Neither result required any new
theory, only a mapping.

Section~\ref{sec:stc} shows where the formalism stops. What has to be treelike
is not the network but the incidence structure of the complexes, so a
representation whose complexes overlap buys nothing: the obvious hypergraph
mapping of a static triadic closure graph is as wrong as heterogeneous mean
field. The same problem mapped onto the backbone, with the occupation state
promoted to a layer, is solved exactly --- closed-form threshold, order parameter
and critical exponents. The failure of mean field reported there
\cite{cirigliano2025} is not a failure of the generating function method but of
applying it to the clustered graph rather than to the object that is treelike.
Finding the representation in which the complexes do not overlap is the work,
and it need not be the obvious one.

\section*{Code availability}

All symbolic calculations reported here, the Monte Carlo simulations used to
validate them, and the scripts that generate every figure are available in the
\code{chygraph} package \cite{chygraphcode}. The methods named throughout the
text and collected in Table~\ref{tab:code} are those of that package, and the
constructors of Fig.~\ref{fig:code} are in its \code{applications} module. The
results reported here are covered by its test suite.

\begin{acknowledgments}
The calculations were assisted by Claude Opus~5, an AI assistant developed by
Anthropic. Nodes \& Links Ltd provided support in the form of salary for Alexei
Vazquez but did not have any additional role in the conceptualization of the
study, the analysis, the decision to publish or the preparation of the
manuscript.
\end{acknowledgments}

\end{document}